\documentstyle[epsfig]{article}

\advance \voffset by -2.2cm
\advance \hoffset by -2cm
\advance \rightmargin by 1 in

\def\sm{Standard Model}
\def\tev{\hbox{TeV}}
\def\gev{\hbox{GeV}}
\def\lcal{{\cal L}}
\def\ocal{{\cal O}}
\def\up#1{^{(#1)}}
\def\su#1{SU(#1)}
\def\ui{U(1)}
\def\gesim{\,{\raise-3pt\hbox{$\sim$}}\!\!\!\!\!{\raise2pt\hbox{$>$}}\,}
\def\lesim{\,{\raise-3pt\hbox{$\sim$}}\!\!\!\!\!{\raise2pt\hbox{$<$}}\,}
\def\etal{{\it et al.}}
\def\ibid{{\it ibid.}}
\def\whatjournal{P}

\newlinechar=`\^^J

\def\ordernpb#1#2#3{{\bf#1} (#3) #2}
\if P\whatjournal {\global\def\order#1#2#3{\orderprd{#1}{#2}{#3}}}
                    \immediate\write16{^^J PRD references ^^J}\else
                   {\global\def\order#1#2#3{\ordernpb{#1}{#2}{#3}}}
                    \immediate\write16{^^J NPB references ^^J}
\fi

\def\ap#1#2#3{{\rm Ann. Phys.\ }\order{#1}{#2}{#3}}

\def\app#1#2#3{{\rm Acta Phys. Pol. {\bf B}}\order{#1}{#2}{#3}}
\def\arnps#1#2#3{{\rm Ann. Rev. Nucl. Phys. Sci. }\order{#1}{#2}{#3}}
\def\ijmpa#1#2#3{{\rm Int. J. of Mod. Phys. {\bf A}}\order{#1}{#2}{#3}}

\def\npb#1#2#3{{\rm Nucl. Phys. {\bf B}}\order{#1}{#2}{#3}}

\def\physa#1#2#3{{\rm Physica {\bf A}}\order{#1}{#2}{#3}}

\def\plb#1#2#3{{\rm Phys. Lett. {\bf B}}\order{#1}{#2}{#3}}

\def\prep#1#2#3{{\rm Phys. Rep.\ }\order{#1}{#2}{#3}}
\def\prl#1#2#3{{\rm Phys. Rev. Lett.\ }\order{#1}{#2}{#3}}

\def\prd#1#2#3{{\rm Phys. Rev. {\bf D}}\order{#1}{#2}{#3}}

\begin{document}
\title{Non-standard t production at the NLC\footnote{Talk prsented in
{\sl Beyond the Standard Model V}, April 29 - May 4, 1997, Kvikne's Hotel,
Balholm, Norway.}}

\author{Jos\'e Wudka\footnote{Work supported in
part through funds provided
by the U.S. Department of Energy under contract FDP-FG03-94ER40837.} \\
\small Department of Physics \\
\small University of California-Riverside \\
\small California, 92521-0413}

\maketitle

\begin{picture}(-20,-50)(0,0)
\put(380,190){{\Large UCRHEP-T192}}
\end{picture}

\begin{abstract}
I present a brief overview of the advantages of using  a consistent
effective Lagrangian approach in parameterizing virtual new physics
effects. I then apply the formalism to certain top-quark processes.
\end{abstract}

\section{Introduction}

In this talk I will discuss the possibilities of detecting non-\sm\
physics thorough virtual effects at a high-energy linear collider.
In the description of these effects I will use the effective Lagrangian
formalism.

\section{Effective lagrangians}

It is a commonly held belief that the Standard Model is but the
low-energy limit of a more fundamental theory. In fact there is a myriad
of models which reduce to the \sm\ at low energies (below,
say, $1\tev$) but which exhibit a plethora of new effects at smaller
scales. I will adopt this paradigm with the added
condition that there is a energy gap between the \sm\ scale $
v \sim 250 \gev $ and the scale of new physics $ \Lambda $. One important
goal of the approach I will follow (for general references see
\cite{basics} and references therein) is to obtain reliable estimates (or
reliable bounds) for $ \Lambda $ using current data; this information can then
be used to estimate the energy at which new colliders must operate.

Note that there are interesting models which do {\em not} satisfy $ v
\ll \Lambda $. For example, many supersymmetric theories predict light,
non-\sm\ scalars of masses below $ 200 \gev $~\cite{susy}. One can discuss
 such theories using the formalism to be developed below, but in order
to do so the spectrum at low energies must be modified to include all
the light supersymmetric particles. I will not consider this possibility
here (see~\cite{piriz.wudka}).

Given the presence of a gap we can imagine integrating out all the heavy
excitations of the theory. The effective interactions (for the light
particles) generated in this manner are summarized in an effective
Lagrangian.

Schematically, denoting the heavy fields
by $ \Phi $, the light fields by $ \phi $, and the action for the theory
underlying the \sm\ by $ S ( \Phi , \phi ) $, then the
effective action is %
\def\seff{S_{\rm eff}}
\begin{equation}
\seff ( \phi )  = - i \ln \left[ \int [ d\Phi ]\; \exp ( i S ) \right]
\end{equation}
Note that $ \seff $ will have a dependence on  $ \Lambda $, and that $
\Lambda $
assumed much larger than any of the light physics scales
(including the energy at the available experiments). Thus one can
do an expansion in powers of $ \Lambda $~\cite{basics}
(up to possible logarithmic factors),
\begin{equation}
\seff = \sum_{ n=-4}^\infty { 1 \over \Lambda^n } \int d^4x \; \lcal_n
\end{equation}
where $ \lcal_n $ can be expanded as a linear combination of {\em
local} operators,
\begin{equation}
\lcal_n = \sum_a f_a\up n \ocal_a \up n
\end{equation}
(any logarithmic dependence on $ \Lambda $ is contained in the
coefficients $ f_a\up n $).

The terms $ \lcal_n , \ n\le 0 $ correspond to the \sm; the $
\Lambda $ dependence of these terms is unobservable~\cite{dec.thm}. The
contributions $ \lcal_n , \ n > 0 $ summarize the virtual heavy-physics
effects. If the action $S$ is known then one can calculate the
coefficients $ f_a\up n $ in terms of the parameters of the heavy
theory, but one can also take a complementary approach and parameterize
{\em all} possible heavy physics effects using these quantities. This
is the approach taken here.

There are no general statements concerning the {\em global} symmetry
properties of the various terms $ \lcal_n $. It is quite possible for
some terms to have a given global symmetry which is absent in others. A
clear example is baryon number violation:  the \sm, corresponding to
$ \lcal_n, \ n \le 0 $, automatically conserves $B$ (ignoring possible
instanton effects~\cite{thooft.inst}); on the other hand there are
contributions
to $ \lcal_2 $ which violate $B$ (for example, assuming the underlying
theory to be the $\su5$ GUT there are well-known baryon-violating
operators of dimension 6 generated by the exchange of heavy
vectors~\cite{su5.b.violation}).

In contrast local symmetries must permeate all the $ \lcal_n
$~\cite{veltman}. Since the \sm\ is assumed to be $ \su3
\times \su2 \times \ui $ symmetric, the same will be true for all
higher-dimensional operators $ \ocal_a\up n $. Since each operator will
then involve several interactions this can result in a reduction in the
number of undetermined parameters.

For example, the dominating
non-standard contributions to the triple {\em and} quartic
gauge-boson vertices (excluding
gluons) appear in $ \lcal_2 $ and involve only 4
independent coefficients. In contrast the most general Lorentz invariant
expression for the triple gauge-boson (not including gluons) vertices
involve 13 unknown coefficients~\footnote{Subdominant contributions
to this vertex, generated by $ \lcal_{ n \ge 3 } $ will generate the
remaining 9 contributions, but the corresponding coefficients are
suppressed by a factor of $ ( v / \Lambda )^k, \ k \ge 2$.}

The effective approach described above is natural~\cite{thooft},
consistent~\cite{basics} and its predictions have been verified
repeatedly (even for strongly coupled theories, see for
example~\cite{examples}).

\section{Segregating operators}

One important feature of the effective Lagrangian approach is the possibility
of
estimating the coefficients $ f_a\up n $. It then becomes possible
to isolate those operators whose coefficients are not {\it a priori}
suppressed and will potentially generate the strongest deviations from
the \sm.

The coefficients estimates strongly depends on the low energy scalar
spectrum. I will consider two possibilities (for a more complicated
scenario see~\cite{perez}):

\begin{itemize}

\item {\sl Light Higgs case:} the light spectrum is taken to correspond to
that of the \sm\ with a single light doublet. In this case, assuming
naturality~\cite{thooft}, the heavy theory should be weakly
coupled~\cite{jw.review}. The dominating operators will then be those
generated at tree level; the list of such operators appeared in
Ref.~\cite{arzt}. Subdominant operators appear with coefficients
suppressed by a factor $ \sim 1/ ( 4 \pi)^2 $.

\item{\sl Chiral case:} the light spectrum corresponds to
that of the \sm\ with no physical scalars. In this case the symmetry
breaking sector is strongly coupled~\cite{georgi} and the coefficients
can be estimated using naive dimensional analysis~\cite{georgi.manohar}.

\end{itemize}

With these estimates $ \Lambda $ has a direct interpretation. For the
light Higgs case it represents the mass of a heavy excitation; for the
chiral case it represents the scale at which the new interactions become
apparent.
In this talk I will consider only the light Higgs case.

\medskip

The above estimates are verified in {\em all} models where calculations
and/or data are available. In choosing processes with which to probe the
physics underlying the \sm\ one should therefore concentrate on
reactions where the dominating operators (those with the largest
coefficients) contribute.

As an example one can study the $WW \gamma $ and $WWZ$ vertices (CP
conserving) in the case where there are light scalars. The dominating
(non-\sm) operators have dimension 6; there are two of them:
\begin{eqnarray}
\ocal_W &=& g^3 \epsilon_{IJK} W^I_\mu{}^\nu W^J_\nu{}^\rho W^K_\rho{}^\mu
\nonumber\\ \ocal_{WB} &=& g g' \left( \phi^\dagger \tau^I \phi\right) W^I_{
\mu \nu} B^{ \mu \nu}
\end{eqnarray}
where $ W^I_{ \mu \nu} $ and $ B_{ \mu \nu} $ denote the $ \su2$ and $
\ui $ field-strengths respectively, with gauge coupling constants $g$
and $ g' $; $I, J,$ etc. denote $ \su2 $ indices and $\phi$ the \sm\
doublet. Gauge coupling constants are explicitly included since gauge
field couple universally. Incidentally it is worth noting that $ \ocal_W
$ is the only CP conserving non-\sm\ operator generating vertices with
$ \ge4 $ (electroweak) gauger bosons.

The effective Lagrangian contains terms $ ( f_W / \Lambda^2 ) \ocal_W +
(f_{ W B} / \Lambda^2 )
\ocal_{WB} $ which in terms of the usual notation~\cite{hikasa.et.al}
translates into
\begin{eqnarray}
&& \lambda_\gamma = \lambda_Z = { 6 m_W^2 g^2 \over\Lambda^2 } f_W ;
\nonumber \\
&& \Delta \kappa_\gamma = \Delta \kappa_Z = { 4 m_W^2 \over \Lambda^2 } f_{
WB}
\end{eqnarray}
Since the operators are loop generated we have $ f_W , f_{WB} \sim 1/ (
16 \pi^2 ) $ and
\begin{equation}
\lambda \sim \left( { 10 \gev \over \Lambda } \right)^2 ; \qquad
\Delta \kappa \sim \left( { 15 \gev \over \Lambda } \right)^2
\end{equation}
so that a measurement stating $ \lambda < 0.05 $ corresponds to $
\Lambda > 45 \gev $. In order to obtain non-trivial information
about $ \Lambda $ from $ \lambda $ or $ \Delta \kappa $
we need to measure these coefficients to a
precision of $ \sim 10^{-4}$.

On the other hand the effective operator $ ( f_{ e e \mu \mu } / \Lambda^2 )
( \bar e_L \gamma^\alpha e_L ) ( \bar \mu_L \gamma^\alpha \mu_L ) $ is
generated at tree level ($ f_{ e e \mu \mu } \sim 1 $) and current
bounds~\cite{4fermion} correspond to $ \Lambda \gesim 0.8 \tev $. This
implies that any new (weakly-coupled) physics generating this operator
will not be seen directly below this scale.

An example may serve to illustrate the above results. The effective
Lagrangian describing neutron $ \beta $ decay is $ \lcal_{n \beta} \sim
G_F \left( \bar n \gamma^\alpha p \right) \left( \bar e \gamma_\alpha \nu
\right) $, with $ G_F \sim ( \hbox{mass})^{-2} $. The above arguments
suggest $ G_F \sim 1/Lambda^2 $ with $ \Lambda $ of the order of the
mass of a heavy particle (up to coupling constants). But one could also
have written $ G_F = \vartheta / m_n^2 $ where $ m_n $ is the neutron
mass, which is certainly fine, one must only remember that $ \vartheta
\sim ( m_n / v)^2 \sim 10^{-5} $.

\medskip

It is, of course, possible for some coefficients to be suppressed by an
unknown symmetry. In this case one cannot distinguish between such a
suppression and a large value of $ \Lambda $~\footnote{Which is not
inconsistent: if new physics with scale $ \Lambda $ does not generate a
given operator $ \ocal $, it is still possible for some heavier physics
with scale $ \Lambda' $ to generate $ \ocal $; the bounds obtained then
refer to $ \Lambda' $.}. In contrast there is no
known mechanism for enhancing the above estimates by more than a factor
$ \lesim 10 $~\cite{einhorn}. For example, if one imagines that there are
$N$ particles contributing to a given operator at the one-loop level the
coefficient corresponding to this operator will be $ \sim N/ ( 4 \pi)^2
$ which can be $ O ( 1 ) $ for $ N = O ( 100 ) $. In this case, however,
the theory cannot be analyzed using perturbation theory, in particular,
the Higgs mass becomes of order $ \Lambda $ and disappears from the
low-energy spectrum~\cite{jw.review}.

\section{Dominating operators involving the top quark}

In the case where there are light scalars there are three types of
operators involving the top quark generated at tree level. All these
operators are $ \su3 \times \su2 \times \ui $ gauge invariant.

\begin{itemize}

\item {\sl Four fermion  interactions.} Examples are,
\begin{eqnarray}
&& \left( \bar t_R \gamma^\mu t_R \right) \left( \bar e_R \gamma_\mu
e_R \right) \nonumber \\
&& \left( \bar \nu_L\up e e_R \right) \left( \bar b_L t_R \right) -
\left( \bar e_L e_R \right) \left( \bar t_L t_R \right)
\label{eq:four.fer}
\end{eqnarray}
(where the second is suppressed since it violates chiral symmetry and
contributes to the electron mass).

\item {\sl Gauge-boson couplings.} For example
\begin{equation}
i \left( \phi^\dagger D_\mu \phi \right) \left( \bar t_R \gamma^\mu t_R
\right) \label{eq:gauge.coupl}
\end{equation}
where $ \phi $ denotes the \sm\ doublet.

\item {\sl Scalar couplings.} For example, in the unitary gauge,
\begin{equation}
H^3 \bar t_L t_R
\end{equation}
where $H$ denotes the physical scalar.

\end{itemize}

In contrast operators such as $ \left( \bar t \sigma_{ \mu \nu} t
\right) F^{ \mu \nu } $ and $ \left( \bar t \gamma_\mu \partial_\nu t
\right) F^{ \mu \nu } $ are generated at one loop by the underlying theory
and their coefficients are suppressed by a factor $ \sim 1/ ( 4 \pi )^2
$.

{}From this one can infer the types of reactions which involve the top
quark and which can best probe the physics underlying the \sm. It is
also possible to determine the {\em type} of new physics which generates
the higher-dimensional operators. For example, the first of the
operators in (\ref{eq:four.fer}) would be generated by a heavy vector, the
operators (\ref{eq:gauge.coupl}) are also generated by virtual heavy
vector bosons~\cite{arzt}, etc. In many reactions
only one operator dominates the cross
section, in these cases one can also specify the type(s) of new physics
which are probed by this process under consideration.

\section{Dominating new physics effects in $t \bar t$ production
through $W$ fusion}

One reaction where new physics effects might be probed is in $t \bar t $
production through $W$ fusion. This process is interesting among other
reasons, because the cross section increases with energy~\footnote{The
corresponding \sm\ calculations can be found in~\cite{wwtt}. A related
calculation at a $ \mu \mu $ collider can be found in~\cite{th}.}.

\begin{figure}[ht]
\centerline{\epsfig{file=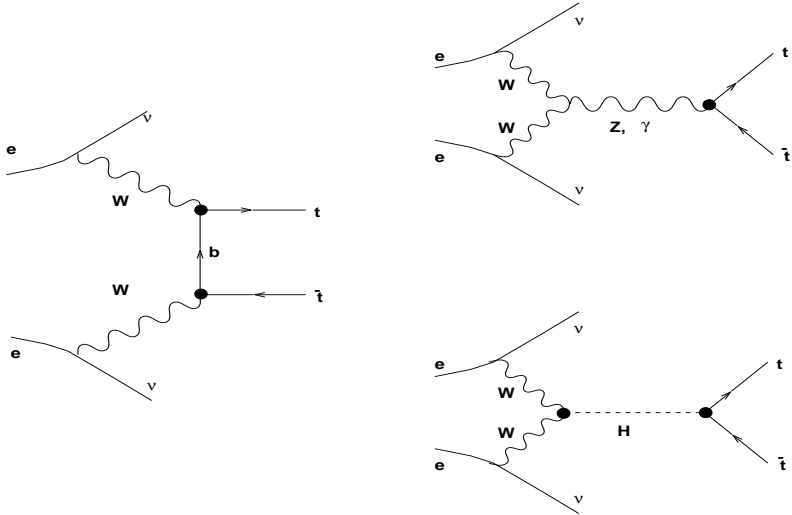, height=300pt,width=4in,
                    bbllx=-3pt, bblly=-270pt, bburx=609pt, bbury=522pt,
                    rheight=200pt}}
\vspace{10pt}
\caption{Graphs for $t \bar t$ production through $W$ fusion. heavy dots denote
vertices affected by the effective operators.}
\label{fig:graphs}
\end{figure}

The new physics effects which can be probed in this process are those
modifying the $ Wtb$, $Ztt$, $WWH$ and $Htt$ vertices. The relevant
operators are  (the contributing graphs are given in fig.~\ref{fig:graphs}).

\medskip

\begin{tabbing}
$ \ocal_\phi\up1 = \left( \phi^\dagger \phi \right) \left[ \left( D_\mu \phi
\right)^\dagger \left( D_\mu \phi \right) \right] $ \= \quad {\it Vertices
affected} \kill
{\it Operator} \> {\it Vertices affected} \\
$ \ocal_{ u \phi } = \left( \phi^\dagger \phi \right) \left( \bar q t_R \tilde
\phi \right) $ \> $ H t \bar t$ \\
$ \ocal_{ \phi q }\up1 = i \left( \phi^\dagger D_\mu \phi \right) \bar q
\gamma^\mu q $ \> $ Z t \bar t$, $ H t \bar t $ \\
$ \ocal_{ \phi q }\up3 = i \left( \phi^\dagger \tau^I D_\mu \phi \right) \bar q
\tau^I \gamma^\mu q $ \> $ Z t \bar t$, $ W t b $, $ H t \bar t$ \\
$ \ocal_{ \phi u } = i \left( \phi^\dagger D_\mu \phi \right) \bar t_R
\gamma^\mu t_R $ \> $ Z t \bar t$, $ H t \bar t $ \\
$ \ocal_{ \phi \phi } = i \left( \phi^T \epsilon D_\mu \phi \right) \bar t_R
\gamma^\mu b_R $ \> $ W t b $ \\
$ \ocal_\phi\up1 = \left( \phi^\dagger \phi \right) \left[ \left( D_\mu \phi
\right)^\dagger \left( D_\mu \phi \right) \right] $ \> $ W W H $ \\
\end{tabbing}

The existing data only bounds the $Wtb$ coupling for which $ \Lambda >
500\gev $ for a left-handed coupling and $ \Lambda > 300\gev $ for a
right-handed one. It is worth pointing out that the $Zt \bar t $
coupling can be measured to a 1\%\ accuracy at both the LHC and the
NLC~\cite{lmy}

\begin{figure}[ht] 
\centerline{\epsfig{file=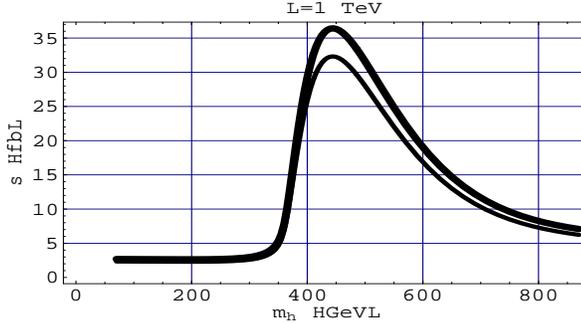, height=300pt,width=4in,
                    bbllx=-3pt, bblly=-270pt, bburx=609pt, bbury=522pt,
                    rheight=100pt}}
\vspace{10pt}
\caption{Total cross section for $ t \bar t $ production through $W$
fusion (calculated using the effective $W$ approximation); \sm: lower
curve, \sm plus effective operators: top curve. The effective operator
coefficients were all chosen t be $ \pm 1$ with the signs chosen to give
the maximum effect.}
\label{fig:total.cs}
\end{figure}

The total cross section proves to be a mediocre probe for new physics;
see for example, fig~\ref{fig:total.cs}

\begin{figure}[ht] 
\centerline{\epsfig{file=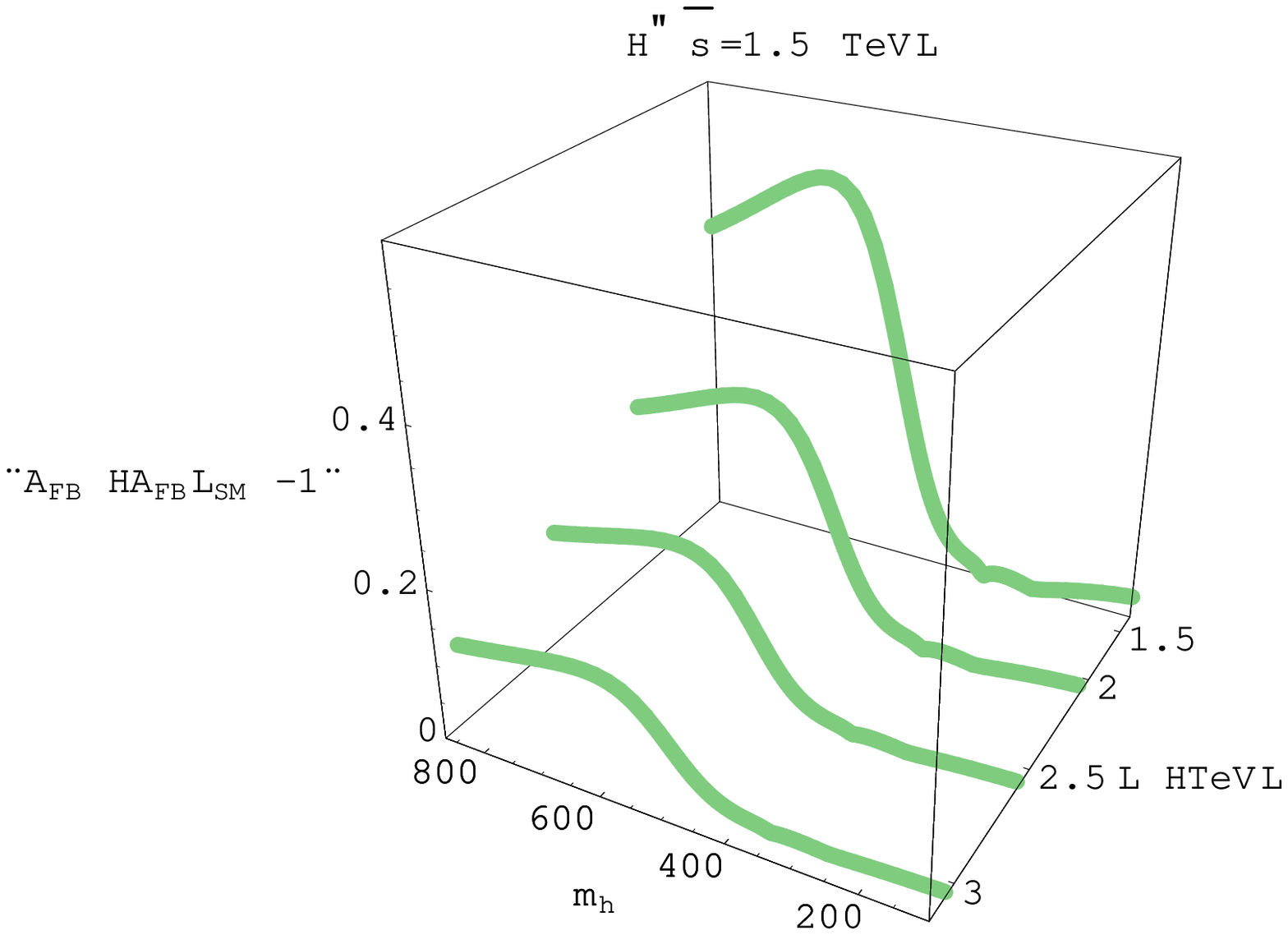, height=400pt,width=4.7in,
                    bbllx=-3pt, bblly=-220pt, bburx=609pt, bbury=572pt,
                    rheight=170pt}}
\vspace{10pt}
\caption{Deviation from the \sm\ in the forward-backward asymmetry in $
t \bar t $ production though $W$ fusion}
\label{fig:afb}
\end{figure}

A more sensitive probe is the forward-backward asymmetry, $ A_{FB}$
which can exhibit deviations of  few$\times 10\%$ from the
\sm\ prediction. Assuming an efficiency of 18\%~\cite{efficiency} we see
from fig.~\ref{fig:afb} that a $ 1.5 \tev $ NLC will be sensitive to
scales $ \Lambda $ up to $ \sim 2.5 \tev $.

\begin{figure}[ht] 
\centerline{\epsfig{file=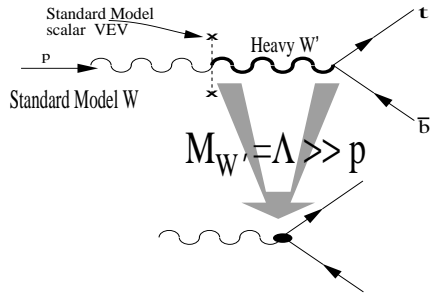, height=250pt,width=3in,
                    bbllx=-3pt, bblly=-70pt, bburx=609pt, bbury=722pt,
                    rheight=100pt}}
\vspace{10pt}
\caption{Possible modification of the $Wtb$ vertex due to the presence
of a heavy $W'$.}
\label{fig:wtb}
\end{figure}

The most important contribution to this process is the $t$-channel $b$
quark exchange and so this process is most sensitive to the $Wtb$
vertex. The deviations from the \sm\ in this interactions come from a
heavy $W'$, as, for example in fig.~\ref{fig:wtb}

\section{Other processes}

\begin{figure}[ht] 
\centerline{\epsfig{file=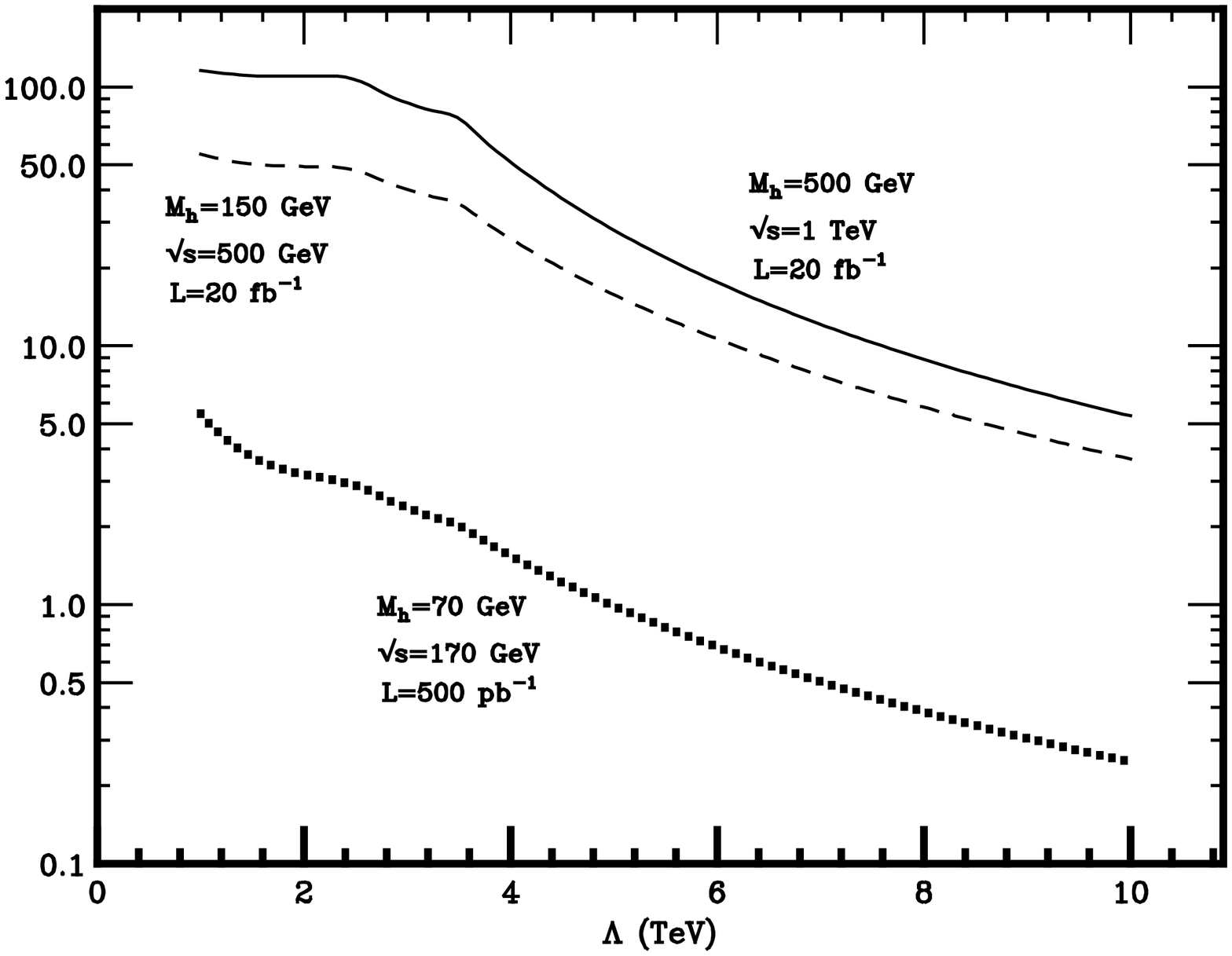, height=3.5in,width=3.5in,
                    bbllx=-3pt, bblly=-120pt, bburx=609pt, bbury=672pt,
                    rheight=2.1in}}
\vspace{10pt}
\caption{Reach in $ \Lambda $ for the process  $ e^+ e^- \rightarrow Z H
$ (including LEP constraints and detection efficiency); $N_{SD}$ denotes
the number of standard deviations from the \sm\ value.}
\label{fig:eezh}
\end{figure}

\begin{itemize}

\item Better probe of new physics involving the top quark
are the reactions $ q q \rightarrow t
\bar t $~\cite{hp} at the Tevatron and
$ e^+ e^- \rightarrow t
\bar t $ at a $ 1 \tev $ NLC~\cite{grzadkowski}. In the first case a
sensitivity to scales up to $ 850 \gev $ can be reached and up to
$ \Lambda \le 5 \tev $ for the second reaction.

\item As an example of reactions which probe physics not containing the
top quark a good example is the process $ e^+ e^- \rightarrow Z H
$~\cite{gw} which is strongly affected by an effective vertex of the form
$ \left( \bar e \gamma_\alpha e \right) Z^\alpha H $ generated by a
heavy $Z'$ vector boson. The reach in $ \Lambda $ of this process is
presented in fig.~\ref{fig:eezh}; for details
see~\cite{gw}~\footnote{This reaction has also been studied in the
context of trip gauge boson vertices and CP violation~\cite{other}.}

\end{itemize}

\section{Conclusions}

The NLC will provide a very powerful tool for probing virtual non-\sm\
physics. The effective Lagrangian approach provides a rationale for the
choice of processes to study, this method has been tested severally and
its predictions and estimates agree with all known calculations and
experiments. Among the processes involving the top quark the
forward-backward asymmetry is sensitive to new physics both in the
direct and the $W$-fusion reactions.


\begin{thebibliography}{99}

\bibitem{basics}
S. Weinberg, \physa{96}{327}{1979}.
J. Polchinski, in: {\sl Recent directions in particle theory: from superstrings
and black holes to the \sm}, Theoretical Advanced Study Institute in High
Elementary Particle Physics (TASI 92), Boulder, Colo., 1-26 Jun, 1992.
Edited by J. Harvey and J. Polchinski. (World Scientific, 1993).
H. Georgi, \arnps{43}{209}{1994}.
%
\bibitem{susy} See, for example, the proceedings of the
DPF/DPB Summer Study on New Directions for High-Energy Physics (Snowmass 96),
Snowmass, CO, 25 Jun - 12 Jul 1996.
S. Dawson, lectures given at {\sl NATO Advanced Study
Institute on Techniques and Concepts of High-energy Physics}, St. Croix, U.S.
Virgin Islands, 10-23 Jul 1996. (hep-ph/9612229)
H. Murayama and M.E. Peskin, \arnps{46}{533}{1996} (hep-ex/9606003).
%
\bibitem{piriz.wudka} D.D. Piriz and J. Wudka, in preparation.
%
\bibitem{dec.thm} T. Appelquist and J. Carazzone, \prd{11}{2856}{1975}.
See also J.C. Collins, {\sl Renormalization} (Cambridge University Press,
1984).
%
\bibitem{thooft.inst} G. 't Hooft, \prl{37}{8}{1976}.
%
\bibitem{su5.b.violation} See, for example, P. Langacker, \prep{72}{185}{1981}
and
references therein.
J. Hisano \etal \npb{402}{46}{1993} (hep-ph/9207279).
%
\bibitem{veltman} M. Veltman, \app{12}{437}{1981}.
%
\bibitem{thooft} G. 't Hooft, in proceedings of the {\sl Cargese Summer
Institute
on Recent Developments in Gauge Theories}, edited by G. 't Hooft \etal (Plenum
Press, 1980)
%
\bibitem{examples} J. Gasser and H. Leutwyler, \npb{250}{465}{1985};
\ap{158}{142}{1984}.
%
\bibitem{perez} J.M. Frere \etal, \plb{292}{348}{1992} (hep-ph/9207258).
M.A. Perez \etal, \prd{52}{494}{1995} (hep-ph/9506457)
%
\bibitem{jw.review} J. Wudka, \ijmpa{9}{2301}{1994} (hep-ph/9406205).
%
\bibitem{arzt} C. Arzt \etal, \npb{433}{41}{1995} (hep-ph/9405214).
%
\bibitem{georgi} See for example, A. Pich, in proceedings of the
{\sl Fifth Mexican School of Particles and Fields}, Guanajuato, Mexico 1992 /
edited by J.L. Lucio M. and M. Vargas (American Institute of Physics, 1994).
(hep-ph/9308351) and references therein.
%
\bibitem{georgi.manohar} A. Manohar and H. Georgi, \npb{234}{189}{1984}.
H. Georgi, \plb{298}{187}{1993} (hep-ph/9207278).
%
\bibitem{hikasa.et.al} K. Hagiwara \etal, \npb{282}{253}{1987}.
%
\bibitem{4fermion} R.M. Barnett \etal, \prd{54}{1}{1996}.
%
\bibitem{einhorn} M.B. Einhorn, in proceedings of the {\sl Workshop on Physics
and Experiments with Linear $e^+ e^- $ Colliders}, Waikoloa,
Hawaii, 26-30 April 1993, edited by  F.A. Harris \etal (World Scientific,
1993).
(hep-ph/9308331); in proceedings of the {\sl International symposium on Unified
symmetry:
in the small and in the large}, January 27-30, 1994 Coral Gables,
Florida, edited by B.N. Kursunoglu \etal.  (Plenum Press, 1995)
(hep-ph/9303323).
%
\bibitem{wwtt} R.P. Kauffman, \prd{41}{3343}{1990}.
M. Gintner and S. Godfrey, in proceedings of the {\sl   OCIP-C-96-5, Jun
1996 DPF / DPB Summer Study on New Directions for High-Energy Physics
(Snowmass 96)}, Snowmass, CO, 25 Jun - 12 Jul 1996 (hep-ph/9612342). T. L.
Barklow, \ibid hep-ph/9704217)
%
\bibitem{th} Muon Quartet Collaboration (V. Barger \etal), in proceedings of
the {\sl Conference on Future High-energy Colliders}, Santa Barbara, CA, 21-25
Oct 1996 (hep-ph/9704290)
%
\bibitem{lmy} F. Larios \etal, \app{27}{3741}{1996} (hep-ph/9609482).
%
\bibitem{efficiency} NLC ZDR Design Group, NLC Physics Working Groups
(S. Kuhlman, \etal) {\sl Physics and Technology of the Next Linear Collider: A
Report
Submitted to Snowmass '96} Report BNL 52-502, FERMILAB-PUB-96/112,
LBNL-PUB-5425,
SLAC-Report-485, UCRL-ID-124160, UC-41 (hep-ex/9605011)
%
\bibitem{hp} C.T. Hill and S.J. Parke, Phys. Rev. D49, 4454 (1994)
(hep-ph/9312324).
%
\bibitem{grzadkowski}B. Grzadkowski , Report IFT-17-95 (hep-ph/9511279).
%
\bibitem{gw}  B. Grzadkowski and J. Wudka, \plb{364}{49}{1995}
(hep-ph/9502415)

\bibitem{other} T. Han and R. Sobey, \prd{52}{6302}{1995} (hep-ph/9507409)
A. Skjold and P. Osland, in proceedings of the {\sl 3rd Tallinn
Symposium on Neutrino Physics}, Lohusalu, Estonia, 7-11 Oct 1995.
(hep-ph/9511453).




\end{thebibliography}
\end{document}